\documentclass[superscriptaddress,twocolumn,showpacs,
amssymb,amsmath,nobibnotes,aps,prd,%
nofootinbib]{revtex4-1}
\pdfoutput=1
\usepackage{graphicx,subfigure,bm,color,psfrag,hyperref}
\usepackage{amsfonts}
\usepackage{lipsum}
\usepackage{mathtools}
\usepackage{verbatim}
\usepackage[normalem]{ulem}
\usepackage[dvipsnames]{xcolor}
\hypersetup{colorlinks,linkcolor={blue},citecolor={red},urlcolor={violet}} 
\begin{document}

\title{Exploring bulk viscous unified scenarios with Gravitational Waves Standard Sirens}

\author{Weiqiang Yang}
\email{d11102004@163.com}
\affiliation{Department of Physics, Liaoning Normal University, Dalian, 116029, P. R. China}

\author{Supriya Pan}
\email{supriya.maths@presiuniv.ac.in}
\affiliation{Department of Mathematics, Presidency University, 86/1 College Street, Kolkata 700073, India}
\affiliation{Institute of Systems Science, Durban University of Technology, PO Box 1334, Durban 4000, Republic of South Africa}

\author{Eleonora Di Valentino}
\email{e.divalentino@sheffield.ac.uk}
\affiliation{School of Mathematics and Statistics, University of Sheffield, Hounsfield Road, Sheffield S3 7RH, United Kingdom}

\author{Celia Escamilla-Rivera}
\email{celia.escamilla@nucleares.unam.mx}
\affiliation{Instituto de Ciencias Nucleares, Universidad Nacional Aut\'onoma de M\'exico, Circuito Exterior C.U., A.P. 70-543, M\'exico D.F. 04510, M\'exico}

\author{Andronikos Paliathanasis}
\email{anpaliat@phys.uoa.gr}
\affiliation{Institute of Systems Science, Durban University of Technology, PO Box 1334, Durban 4000, Republic of South Africa}
\affiliation{Departamento de Matem\'{a}ticas, Universidad Cat\'{o}lica del Norte, Avda. Angamos 0610, 1280 Casilla, Antofagasta, Chile}
\affiliation{Mathematical Physics and Computational Statistics Research Laboratory, Department of Environment, Ionian University, Zakinthos 29100, Greece}
%---------------------------------------------------

%%%%%%%%%%%%%%%%%%%%%%%%%%%%%%%%%%%%%%%%%%%%%%%%%%%
\begin{abstract}
We consider the unified bulk viscous scenarios and constrain them using the Cosmic Microwave Background observations from Planck 2018 and the Pantheon sample from Type Ia Supernovae. Then we generate the luminosity distance measurements from ${\cal O}(10^3)$ mock Gravitational Wave Standard Sirens (GWSS) events 
for the proposed Einstein Telescope. We then combine these mock luminosity distance measurements from the GWSS with the current cosmological probes in order to forecast how the mock GWSS data could be effective in constraining these bulk viscous scenarios. Our results show that a non-zero time dependent bulk viscosity in the universe sector is strongly preferred by the current cosmological probes and will possibly be confirmed at many standard deviations by the future GWSS measurements. We further mention that the addition of GWSS data can significantly reduce the uncertainties of the key cosmological parameters obtained from the usual cosmological probes employed in this work.
\end{abstract}

\maketitle

%\pacs{98.80.-k, 95.35.+d, 95.36.+x}
%----------------------------------------------

%----------------------------------------------
%%%%%%%%%%%%%%%%%%%%%%%%%%%%%%%%%%%%%%%%%%%%%%%%%%%
\section{Introduction}

Understanding the nature of dark matter and dark energy has been a challenge for cosmologists. The standard cosmological model, namely, the so-called $\Lambda$-Cold Dark Matter ($\Lambda$CDM) model representing a mixture of two non-interacting fluids $-$ a positive cosmological constant ($\Lambda >0$) and a cold dark matter component, has undoubtedly proved its unprecedented success by explaining a large span of astronomical data. However, this simplest cosmological scenario has some limitations. For example,  the cosmological constant problem~\cite{Weinberg:1988cp} and the coincidence problem~\cite{Zlatev:1998tr} have already questioned the existing assumptions in the $\Lambda$CDM model, e.g.~constant energy density of the vacuum  and the non-interacting nature between $\Lambda$ and CDM. These limitations motivated the cosmologists to find alternative cosmological scenarios beyond $\Lambda$CDM by relaxing the above assumptions, and as a consequence, several new cosmological models were introduced, see~\cite{Copeland:2006wr,Nojiri:2006ri,Sotiriou:2008rp,DeFelice:2010aj,Capozziello:2011et,Clifton:2011jh,Bamba:2012cp,Cai:2015emx,Nojiri:2017ncd,Bahamonde:2021gfp} for a review of various dark energy and modified gravity models. Additionally, the appearance of cosmological tensions at many standard deviations between Planck~\cite{Aghanim:2018eyx} (assuming $\Lambda$CDM in the background) and other cosmological probes, such as distance ladders~\cite{Wong:2019kwg,Riess:2020fzl,Shajib:2019toy,Birrer:2020tax,Freedman:2021ahq,Anand:2021sum,Blakeslee:2021rqi,Pesce:2020xfe,Shah:2021onj,deJaeger:2022lit,Riess:2021jrx} or weak lensing~\cite{Heymans:2012gg,Hildebrandt:2016iqg,KiDS:2020suj,DES:2017myr,DES:2021wwk} and galaxy cluster data~\cite{Planck:2015lwi,SPT:2018njh,Hasselfield:2013wf} has further  weakened the confidence in the $\Lambda$CDM cosmological model~\cite{Riess:2020sih,Verde:2019ivm,DiValentino:2020vnx,DiValentino:2020vvd,DiValentino:2020zio}. Thus, the list of cosmological models aiming to address the cosmological tensions is increasing in time, see the review articles~\cite{Knox:2019rjx,Jedamzik:2020zmd,DiValentino:2021izs,Perivolaropoulos:2021jda,Schoneberg:2021qvd,Abdalla:2022yfr,Kamionkowski:2022pkx} and references therein. Given the fact that the origin of dark matter and dark energy is not clearly understood yet, thus,  there is no reason to favor any particular cosmological theory over others. As a result, various ways have been proposed to interpret the dynamics of the dark sector in terms of dark matter and dark energy. The simplest  assumption is the consideration of independent evolution of these dark fluids. The generalization of the above consideration is the assumption of a non-gravitational interaction between these dark sectors. On the other hand, a heuristic approach is to consider a unified dark fluid that can explain  the dynamics of dark energy and dark matter at cosmological scales.  
The attempt to unify the dark sector of the Universe began long back ago. The most simplest unified dark sector models can be constructed in the context of Einstein gravity with the introduction of a generalized equation of state $p = \mathcal{F}(\rho)$, where $p$ and $\rho$ are respectively the pressure and energy density of the unified dark sector and $\mathcal{F}$ is an analytic function of the energy density, $\rho$. The well known unified cosmological models, such as the Chaplygin gas model~\cite{Chaplygin} and its successive generalizations, namely, the generalized Chaplygin gas, modified Chaplygin gas, see Refs.~\cite{Kamenshchik:2001cp,Bilic:2001cg,Gorini:2002kf,Gorini:2005nw,Bento:2002ps,Xu:2010zzb,Lu:2009zzf,Benaoum:2002zs,Debnath:2004cd,Lu:2008zzb,Xu:2012ca,Escamilla-Rivera:2019hqt} and some other unified cosmological scenarios as well~\cite{Hova:2010na,Hernandez-Almada:2018osh,Yang:2019jwn} belong to this classification. While it is essential to mention that a subset of the unified models has been diagnosed with exponential blowup in the matter power spectrum which is not consistent with the observations~\cite{Sandvik:2002jz}, however, this does not rule out the possibility of unified models aiming to cover a wide region of the universe evolution because a new kind of unified fluid may avoid such unphysical activities. 
The unified cosmological models can also be developed by considering a relation like $p  = \mathcal{G} (H)$ where $\mathcal{G}$ is an analytic function of $H$, the Hubble function of the Friedmann-Lema\^{i}tre-Robertson-Walker (FLRW) line element. Apparently, theories with $p = \mathcal{F}(\rho)$ and $p  = \mathcal{G} (H)$ seem identical, however, this is only true in spatially flat FLRW universe. For a curved universe, the two approaches are not the same.

In the present work we are interested to study a particular class of unified models endowed with bulk viscosity. The cosmological fluids allowing bulk viscosity as an extra ingredient can explain the accelerating expansion of the universe, and hence they are also enlisted as possible alternatives to the standard $\Lambda$CDM cosmology in the literature~\cite{Yadav:2021hqd,Normann:2021bjy}.
Following an earlier work Ref.~\cite{Yang:2019qza} where an evidence of non-zero bulk viscosity was preferred by the current cosmological probes, in the present article, we use the  simulated Gravitational Waves Standard Sirens (GWSS) measurements from the Einstein Telescope~\cite{Maggiore:2019uih}\footnote{https://www.einsteintelescope.nl/en/}  in order to quantify the improvements of the cosmological parameters, if any, from the future GWSS measurements. As the gravitational waves (GW) have opened a new window for astrophysics and cosmology, therefore, it will be interesting to investigate the contribution from the simulated GWSS data, once combined with the current cosmological probes. This motivated many investigators to use  the mock GWSS data matching the expected sensitivity of the Einstein Telescope  to constrain a class of cosmological models, see for instance,~\cite{Zhao:2010sz,Cai:2016sby,Zhang:2017sym,Cai:2017aea,Yang:2019bpr,Yang:2019vni,Bachega:2019fki,Zhou:2019gda,Yang:2020wby,Mitra:2020vzq,Escamilla-Rivera:2020llu,Pan:2021tpk}.
In particular, the combined analysis of simulated GWSS measurements from Einstein Telescope and the standard cosmological probes has proven to be very effective for a class of cosmological models, in the sense that the error bars in the key cosmological parameters of these cosmological models are significantly reduced thanks to the mock GWSS dataset~\cite{Zhang:2018byx,Wang:2018lun,DiValentino:2018jbh,Du:2018tia,Yang:2019bpr,Yang:2019vni,Yang:2020wby}, however,  
in some  specific $f(R)$ theories of gravity, the generated mock GWSS from the Einstein Telescope may not be very much helpful to give stringent  constraints on them during its first phase of running ~\cite{Matos:2021qne}.  Thus, one may expect that the constraining power of the Einstein Telescope may depend on the underlying cosmological model. 
Aside from the future GWSS measurements from the Einstein Telescope,  one can also use the simulated GWSS measurements from other GW observatories, such as, Laser Interferometer Space Antenna (LISA)~\cite{Allahyari:2021enz,Corman:2020pyr,LISACosmologyWorkingGroup:2019mwx,Cai:2017yww} and DECi-heltz Interferometer Gravitational wave Observatory (DECIGO)~\cite{Hou:2021nxn,Zheng:2020tau}, TianQin~\cite{TianQin:2015yph}. In this article, we  focus only on the simulated GWSS data from Einstein Telescope to constrain the bulk viscous unified scenario.

The paper has been organized as follows: in Sec.~\ref{sec-efe} we discuss the gravitational equations for the bulk viscous scenario. Sec.~\ref{sec-data} describes the observational data that we have considered for the analysis in this work. Sec.~\ref{sec:method} presents the observational constraints on the bulk viscous models, and mainly we discuss how the inclusion of gravitational waves data from the Einstein Telescope improves the constraints. Finally, in Sec.~\ref{sec-conclu} we present the conclusions.

%%%%%%%%%%%%%%%%%%%%%%%%%%%%%%%%%%%%%%%%%%%%%%%%%%%%%%%%%%
\section{Revisiting the bulk viscous scenarios: Background and perturbations}
\label{sec-efe}

As usual, we consider the homogeneous and isotropic space time described by the 
Friedmann-Lema\^{i}tre-Robertson-Walker (FLRW) line element 
\begin{equation}
ds^2 = -dt^2 + a^2 (t) \left[\frac{dr^2}{1-kr^2} + r^2 \left(d \theta^2 + \sin^2 \theta d \phi^2\right)\right],
\end{equation}
where $a(t)$ is the expansion scale factor and $k$ denotes the spatial curvature of the universe. For $k =0,-1, +1$, we have three different geometries of the universe, namely,
spatially flat, open and closed, respectively. In this paper we restrict ourselves to the spatially flat scenario where we assume that (i)~the gravitational sector is described by the Einstein's gravity, (ii)~the matter sector of the universe consists of the relativistic radiation, non-relativistic baryons and a unified bulk viscous fluid which combines the effects of dark matter and dark energy, (iii)~all the fluids are non-interacting with each other. 
Within this framework, we can write down the gravitational field equations as follows (in the units where $8 \pi G = 1$)
\begin{eqnarray}
&&H^2 =  \frac{1}{3}\rho_{\rm tot},\label{efe1}\\
&& 2 \dot{H} + 3 H^2 = - \; p_{\rm tot},  \label{efe2}
\end{eqnarray}
where an overhead dot indicates the derivative with respect to the cosmic time $t$; $H \equiv \dot{a}/a$ is the Hubble expansion rate; $(\rho_{\rm tot}, p_{\rm tot}) = (\rho_r + \rho_b + \rho_u, p_r + p_b + p_{u})$ are the total energy density and total pressure of the cosmic components in which $(\rho_r, p_r)$, $(\rho_b, p_b)$, $(\rho_u, p_u)$ are the energy density and pressure of radiation, baryons and the unified fluid, respectively. The conservation equation for each fluid follows the usual law $\dot{\rho}_i + 3 H (1+w_i)\rho_i = 0$, where the subscript $i$ refers to radiation $(i =r)$, baryons $(i=b)$ and the unified fluid $(i =u)$ and $w_i$ are the standard barotropic state parameters: $w_r = p_r/\rho_{r} = 1/3$, $w_b = p_b/\rho_b =0$ and $w_{u} = p_{u}/\rho_{u} = (\gamma -1)$,  where $\gamma $ is a constant parameter. In general for different values of $\gamma$, say for instance, $\gamma = 0$, we realize a cosmological constant-like fluid endowed with the bulk viscosity and similarly $\gamma = 1$ results in  a dust-like fluid endowed with the bulk viscosity. As the nature of the fluid is not clearly understood and as the observational data play an effective role to understand this nature, thus, in order to be more transparent in this direction we consider  $\gamma$ lying in the interval $[-3, 3]$ which includes both exotic ($ p_u/\rho_u = (\gamma -1) < -1/3 $) and non-exotic 
($ p_u/\rho_u = (\gamma -1) > -1/3$ ) fluids. 
As already mentioned, since the unified fluid has a bulk viscosity, therefore, it enjoys an effective pressure~\cite{Barrow:1988yc}: $p_{\rm eff} = p_{u} - u^{\nu}_{;\nu} \eta (\rho_{u})$, where $u^{\mu}_{; \mu}$ is the expansion scalar of this fluid and  $\eta (\rho_{u})>0$ is the coefficient of the bulk viscosity. 
Thus, in the FLRW background, the effective pressure of the bulk viscous fluid reduces to 
\begin{eqnarray}\label{eff-pressure}
p_{\rm eff} = p_u - 3 H \eta (\rho_u). 
\end{eqnarray}
Since there is no unique selection for the bulk viscous coefficient, $\eta (\rho_{u})$, therefore, we consider a well known choice for it in which the bulk viscous coefficient has a power law evolution of the form~\cite{Barrow:1986yf,Barrow:1988yc,Barrow:1990vx}: 
\begin{eqnarray}\label{BVF}
\eta (\rho_u)=\alpha\rho_{u}^m,
\end{eqnarray}
where $\alpha $ is a positive constant and $m$ is any real number. Notice that for the case $m =0$ we recover the scenario with a constant bulk viscous coefficient. 
Now, with the consideration of the bulk viscous coefficient in (\ref{BVF}), the effective pressure of the unified fluid can be expressed as 
\begin{eqnarray}\label{eq3}
p_{\rm eff}=(\gamma-1)\rho_u-\sqrt{3}\alpha\rho_{\rm tot}^{1/2}\rho_u^m,
\end{eqnarray}
and consequently, one can define the effective equation of state of the viscous dark fluid as 
\begin{eqnarray}\label{eq3}
w_{\rm eff}= \frac{p_{\rm eff}}{\rho_{u}}  =(\gamma-1)-\sqrt{3}\alpha\rho_{\rm tot}^{1/2}\rho_u^{m-1}.
\end{eqnarray}
The adiabatic sound speed for the viscous fluid is given by
\begin{equation}
c_{a,\rm eff}^{2}=\frac{p_{\rm eff}^{\prime}}{\rho_u^{\prime}}=w_{\rm eff}+\frac{w_{\rm eff}^{\prime}}{3\mathcal{H}(1+w_{\rm eff})}.
\end{equation}
where the prime denotes the derivative with respect to the conformal time $\tau$ and  $\mathcal{H}$ is the conformal Hubble parameter, $\mathcal{H}=aH$.
Note that depending on the nature of $w_{\rm eff}$, $c_{a,\rm eff}^{2}$ could be negative, and hence $c_{a,\rm eff}$ could be an imaginary quantity.
This may invite instabilities in the perturbations. Thus, in order to avoid this possible unphysical situation, we consider the entropy perturbations (non-adiabatic perturbations) in the unified dark fluid following the analysis of generalized dark matter~\cite{Hu:1998kj}.

Now we focus on the evolution of the unified bulk viscous fluid at the level of perturbations. In the entropy perturbation mode, the true pressure perturbation comes from the effective pressure given by 
\begin{eqnarray}
\delta p_{\rm eff}&=&\delta p_u-\delta\eta(\nabla_{\sigma}u^{\sigma})-\eta(\delta\nabla_{\sigma}u^{\sigma})\notag\\
&=&\delta
p_u-3H\delta\eta-\frac{\eta}{a}\left(\theta+\frac{h'}{2}\right).
\end{eqnarray}

The effective sound speed of viscous dark fluid for the bulk viscous coefficient (\ref{BVF}) can be defined as 
\begin{eqnarray}
&&c_{s,\rm eff}^{2} \equiv \left(\frac{\delta p_{\rm eff}}{\delta \rho_u}\right)_{rf} \nonumber \\
&&=c_{s}^{2}-\sqrt{3}\alpha m\rho_{\rm tot}^{1/2}\rho_u^{m-1}-\frac{\alpha\rho_u^{m-1}}{a \delta_u}\left(\theta+\frac{h^{\prime}}{2}\right),
\end{eqnarray}
where '$|_{rf}$' denotes the rest frame.  Following the analysis in~\cite{Hu:1998kj}, the sound speed in the rest frame is assumed to be zero, i.e. $c^2_s=0$.

The density perturbation and the velocity perturbation can also be written as~\cite{Hu:1998kj}
%-----------------------------------------------------
\begin{eqnarray}
\delta_u^{\prime}&=&-(1+w_{\rm eff})(\theta_u+\frac{h^{\prime}}{2})+\frac{w_{\rm eff}^{\prime}}{1+w_{\rm eff}}\delta_u \notag\\
&-&3\mathcal{H}(c_{s,\rm eff}^{2}-c_{a, \rm eff}^{2})\left[\delta_u+3\mathcal{H}(1+w_{\rm eff})\frac{\theta_D}{k^{2}}\right], ~\quad
\\
\theta_u^{\prime}&=&-\mathcal{H}(1-3c_{s,\rm eff}^{2})\theta_u+\frac{c_{s,\rm eff}^{2}}{1+w_{\rm eff}}k^{2}\delta_u,
\end{eqnarray}
%---------------------------------------------------
Thus, following the evolution at the background and perturbation level prescribed above, one can now be able to understand the dynamics of the bulk viscous fluid. In this work  
we consider two different bulk viscous scenarios characterized as follows: the bulk viscous model 1 (labeled as BVF1) where we consider $\gamma =1$, and the bulk viscous model 2 (labeled as BVF2) where we keep $\gamma$ as a free parameter. The common parameters in both BVF1 and BVF2 are $\alpha$ and $m$.

%%%%%%%%%%%%%%%%%%%%%%%%%%%%%%%%%%%%%%%%%%%%%%%%%%%%%%%%%%%
\section{Standard Cosmological probes, simulated GWSS data, and methodology}
\label{sec-data}

In this section we describe the cosmological data sets employed to perform the statistical analyses of the present bulk viscous scenarios.  

\begin{itemize}

\item {\bf Cosmic Microwave Background (CMB):} We use the CMB data from the Planck 2018 data release. Precisely, we use the CMB temperature and polarization angular power spectra {\it plikTTTEEE+lowl+lowE}~\cite{Aghanim:2018oex,Aghanim:2019ame}.

\item {\bf Pantheon sample from Type Ia Supernovae (SNIa) data:} Type Ia Supernovae  are the first astronomical data that probed the accelerating expansion of the universe and hence indicated the existence of an exotic fluid with negative pressure (dark energy). Here we use the Pantheon compilation of the SNIa data comprising 1048 data points spanned in the redshift interval $[0.01, 2.3]$~\cite{Scolnic:2017caz}.

\item {\bf Gravitational Waves Standard Sirens (GWSS):} 
We take the mock Gravitational Waves Standard Sirens (GWSS) data generated by matching the expected sensitivity of Einstein Telescope in order to understand the constraining power of the future GWSS data from the Einstein Telescope. The Einstein Telescope is a proposed ground based third-generation (3G) GW detector. The telescope will take a triangular shape and its each arm length will be increased to 10 km, compared to 3 km arm length VIRGO and 4 km arm length LIGO~\cite{Sathyaprakash:2012jk,Maggiore:2019uih}. Thus, due to such increased arm length, the Einstein Telescope will be a potential GW detector by reducing all displacement noises~\cite{Sathyaprakash:2012jk,Maggiore:2019uih}. It is expected that after 10 years of operation, Einstein Telescope will detect $\mathcal{O} (10^3)$ GWSS events. Although the detection of $\mathcal{O} (10^3)$ GWSS events is very optimistic while the number of detections could be low in reality~\cite{Maggiore:2019uih}. As argued in~\cite{Maggiore:2019uih}, the Einstein Telescope will likely to detect 20 $-$ 50 events per year, i.e. 200 $-$ 500 events in 10 years. However, following the earlier works~\cite{Zhao:2010sz,Cai:2016sby,Wang:2018lun,Du:2018tia,Yang:2019vni,Yang:2019bpr,Yang:2020wby,Matos:2021qne,Pan:2021tpk},  in this article, we restrict ourselves to the detection of  $\mathcal{O} (10^3)$ GWSS events by the Einstein Telescope to constrain the bulk viscous scenarios. 
For more features of the Einstein Telescope we refer the readers to~\cite{Maggiore:2019uih}.

We originally generate the mock GWSS luminosity distance measurements  matching the expected sensitivity of the Einstein Telescope after 10 years of full operation. Specifically we create 1000 triples ($z_i$, $d_{L} (z_i)$, $\sigma_i$) where $z_i$ is the redshift of a GW source, $d_L (z_i)$ is the measured luminosity distance at redshift $z_i$ and $\sigma_i$ is the uncertainty associated with the luminosity distance $d_L (z_i)$.  Let us briefly summarize the procedure of generating the mock GWSS dataset and we refer to Refs.~\cite{Du:2018tia,Yang:2019bpr,Yang:2019vni} for more technical details. 
The initial step for generating the mock GWSS dataset is to identify the expected GW sources. We consider the GW events originating from two distinct binary systems, namely, (i) a combination of
a Black Hole (BH) and a Neutron Star (NS) merger, identified as
BHNS and (ii) the binary neutron star (BNS) merger. 
Following the mass distributions as described in Ref.~\cite{Du:2018tia}, the ratio of the number of GW events for the BHNS merger versus BNS merger is taken to be $0.03$ as predicted for the Advanced LIGO-VIRGO
network~\cite{LIGOScientific:2010weo}.  We then determine the merger rate $R(z)$ of sources and from the merger rate of the sources, we determine the  redshift distribution of the  sources, $P (z)$  given by~\cite{Sathyaprakash:2009xt,Zhao:2010sz,Cai:2016sby,Wang:2018lun,Du:2018tia,Yang:2019bpr}

\begin{eqnarray}
    P (z) \propto \frac{4 \pi d_C^2 (z) R (z)}{H (z) (1+z)},
\end{eqnarray}
where $d_C (z) \equiv \int_{0}^{z} H^{-1} (z^\prime) dz^\prime$ is the co-moving distance and for $R (z)$  we take the following piece-wise linear function estimated in~\cite{Schneider:2000sg} (also see~\cite{Cutler:2009qv,Zhao:2010sz,Cai:2016sby,Wang:2018lun,Du:2018tia,Yang:2019bpr}):  $R(z) = 1+2 z$ for $z\leq 1$, $R (z) = \frac{3}{4}(5-z)$,  for $1<z<5$ and $R (z) = 0$ for $z > 5$.  After having $P (z)$, we sample 1000
values of redshifts from this distribution  which represent the redshifts $z_i$ of our 1000 mock GWSS data.

The next step is to choose a fiducial model because while going from the merger rate to the redshift distributions, a fiducial cosmological model is needed since
the expression for $P (z)$ includes both the comoving distance and expansion rate at redshift $z$, i.e. $d_L (z)$ and $H(z)$
respectively.  This $H (z)$ corresponds to the fiducial model. 
As in this article we are interested to investigate how the inclusion of GWSS data improves the constraints on the BVF models,
therefore, we generate two different mock GWSS datasets choosing
BVF1 and BVF2 as the fiducial models. 
We take the fiducial values of the cosmological parameters given by the best fit values of the same cosmological parameters of the BVF1 and BVF2 models obtained from the CMB+Pantheon data analysis.  Now, for the chosen fiducial model(s), one can now estimate the luminosity distance at the redshift $z_i$ using the relation 

\begin{eqnarray}
d_L(z_i) = (1+z_i)\int_0^{z_i}\frac{dz'}{H(z')}\,.
\label{eq:luminosity}
\end{eqnarray}
Thus, after having the luminosity distance $d_{L} (z_i)$ of the GW source, our last job is now to determine the uncertainty $\sigma_i$ associated with this luminosity distance. The determination of the uncertainty $\sigma_i$ directly connects to the waveform of GW because the GW amplitude depends on the luminosity distance (also on the so-called chirp mass~\cite{Zhao:2010sz,Cai:2016sby,Wang:2018lun}) and hence one can extract the information about $d_L (z_i)$ and $\sigma_i$. We refer to Refs.~\cite{Zhao:2010sz,Cai:2016sby,Wang:2018lun,Du:2018tia,Yang:2019bpr} for the technical details to calculate the uncertainties on the luminosity distance measurements. 
The luminosity distance measurement $d_L (z_i)$ has two kind of uncertainties, one is the instrumental uncertainty $\sigma_i^{\rm inst}$  and the other one is the weak lensing uncertainty $\sigma_i^{\rm lens}$. The instrumental error can be derived to be $\sigma_i^{\rm inst} ~(\simeq 2 d_L (z_i)/\mathcal{S}$ where $\mathcal{S}$  is the combined signal-to-noise ratio of the Einstein Telescope)  using the Fisher matrix approach and assuming that the uncertainty on $d_L (z_i)$ is not correlated with the uncertainties on the remaining GW parameters   (see~\cite{Zhao:2010sz,Cai:2016sby,Wang:2018lun,Du:2018tia,Yang:2019bpr}) and the lensing error is  $\sigma_i^{\rm lens} \simeq 0.05 z_i d_L (z_i)$~\cite{Zhao:2010sz}. Thus, the total uncertainty due to the instrumental and the weak lensing uncertainties on $d_L (z_i)$ is  $\sigma_i = \sqrt{(\sigma_i^{\rm inst})^2 + (\sigma_i^{\rm lens})^2}$.  Finally, let us note that the combined signal-to-noise ratio of the GW detector is a very crucial quantity in this context since for the Einstein Telescope, the combined signal-to-noise ratio should be at least $8$ for a GW detection~\cite{Sathyaprakash:2009xt}. Thus, in summary, we generate 1000 GW sources up to redshift $z =2$ with $\mathcal{S} > 8$. For more technical details we refer the readers to Refs.~\cite{Sathyaprakash:2009xt,Zhao:2010sz,Cai:2016sby,Wang:2018lun,Du:2018tia,Yang:2019bpr}.

\end{itemize}

%----------------------------------------------------------------
To constrain the BVF scenarios we modify the publicly available \texttt{CosmoMC} package~\cite{Lewis:2002ah} which is an excellent cosmological code supporting the Planck 2018 likelihood~\cite{Aghanim:2019ame} and it has a convergence diagnostic following the Gelman-Rubin statistic $R-1$~\cite{Gelman:1992zz}.
It is essential to mention that for both BVF1 and BVF2 scenarios, we have used the dimensionless quantity $\beta=\alpha H_0 \rho_{\rm tot,0}^{m-1}$ where $\rho_{\rm tot,0}$ is the present value of $\rho_{\rm tot}$. We further mention here that $\beta = 0$ (equivalently, $\alpha =0$) implies no viscosity and hence the overall picture behaves like a unified cosmic fluid without bulk viscosity.   Thus, in summary, the parameter space of BVF1 and BVF2 are as below: 
\begin{eqnarray}
&&\mathcal{P}_{\rm BVF1} \equiv  \{\Omega_b h^2, 100\theta_{\rm MC}, \tau, n_s, {\rm{ln}}(10^{10} A_s), \beta, m\}\nonumber  \\
&&\mathcal{P}_{\rm BVF2} =\{\Omega_b h^2, 100\theta_{\rm MC}, \tau, n_s, {\rm{ln}}(10^{10} A_s), \beta, m, \gamma\}\nonumber
\end{eqnarray} 
where the description of the free parameters are as follows: $\Omega_bh^2$ is the baryons density, $100 \theta_{MC}$ is the ratio of the sound horizon to the angular diameter distance; $\tau$ is the optical depth, $n_s$ is the scalar spectral index, $A_s$ is the amplitude of the initial power spectrum. 
The flat priors on both cosmological scenarios are shown in Table~\ref{tab-priors}. 
%%%%%%%%%%%%%%%%%%%%%%%%%%%%%%%%%%%%%%%%%%%%%%%%%%%

\begin{table}
\begin{center}
\begin{tabular}{ccccc}
\hline
Parameter       & Priors (BVF1) & Priors (BVF2)\\
\hline 
$\Omega_{b} h^2$     & $[0.005,0.1]$ & $[0.005,0.1]$\\
$\tau$              & $[0.01,0.8]$ & $[0.01,0.8]$\\
$n_s$                        & $[0.5, 1.5]$ & $[0.5, 1.5]$\\
${\rm{ln}}(10^{10}A_{s})$    & $[2.4,4]$ & $[2.4,4]$ \\
$100\theta_{\rm MC}$         & $[0.5,10]$ & $[0.5,10]$\\ 
$\beta$                      & $[0, 1]$ & $[0, 1]$\\ 
$m$                          & $[-2, 0.5]$  & $[-2, 0.5]$\\
$\gamma$                     & $-$  & $[-3, 3]$  \\
\hline
\end{tabular}
\end{center}
\caption{We show the flat priors on all the free parameters associated with the bulk viscous models. }
\label{tab-priors}
\end{table}

%%%%%%%%%%%%%%%%%%%%%%%%%%%%%%%%%%%%%%%%%%%%%%%%%%%
\begingroup                                             
%\squeezetable                                                                                                  
\begin{center}                                                                                                      
\begin{table*}                                                                                                                   
\begin{tabular}{cccccccccc}                                                                                                          
\hline\hline                                                                                                                    
Parameters  & CMB+Pantheon &  CMB+Pantheon+GWSS\\ \hline

$\Omega_b h^2$    & $    0.02232_{-    0.00015-    0.00028}^{+    0.00015+    0.00029}$  &   $    0.02253_{-    0.00014-    0.00026}^{+    0.00014+    0.00028}$ \\

$100\theta_{\rm MC}$   & $    1.02780_{-    0.00055-    0.0011}^{+    0.00058+    0.0011}$ &  $    1.02808_{-    0.00038-    0.00073}^{+    0.00037+    0.00073}$ \\

$\tau$  & $    0.0537_{-    0.0075-    0.015}^{+    0.0074+    0.016}$  &  $    0.0567_{-    0.0078-    0.015}^{+    0.0079+    0.016}$ \\

$n_s$  & $    0.9641_{-    0.0043-    0.0084}^{+    0.0043+    0.0086}$  &  $    0.9686_{-    0.0040-    0.0080}^{+    0.0041+    0.0080}$  \\

${\rm{ln}}(10^{10} A_s)$   & $    3.046_{-    0.015-    0.031}^{+    0.016+    0.031}$  &  $    3.048_{-    0.016-    0.033}^{+    0.016+    0.033}$ \\

$\beta$ &  $    0.430_{-    0.016-    0.034}^{+    0.017+    0.033}$   &  $   0.4262_{-    0.0078-    0.015}^{+    0.0079+    0.016}$  \\

$m$  & $   -0.557_{-    0.059-    0.13}^{+    0.068+    0.12}$ & $   -0.519_{-    0.035-    0.075}^{+    0.038+    0.074}$  \\

$H_0$   & $   68.1_{-    1.1-    2.3}^{+    1.2+    2.2}$  & $   68.30_{-    0.45-    0.85}^{+    0.46+    0.91}$ \\

\hline\hline      
\end{tabular}                                           
\caption{We report the observational constraints on the BVF1 scenario at 68\% and 95\% CL for CMB+Pantheon and CMB+Pantheon+GWSS datasets. }
\label{tab:results-BVF1}                                    
\end{table*}                                                
\end{center}                                                 
\endgroup   
%%%%%%%%%%%%%%%%%%%%%%%%%%%%%%%%%%%%%%%%%%%%%%%%%%%%%%%
\begin{figure*}
\centering
\includegraphics[width=0.7\textwidth]{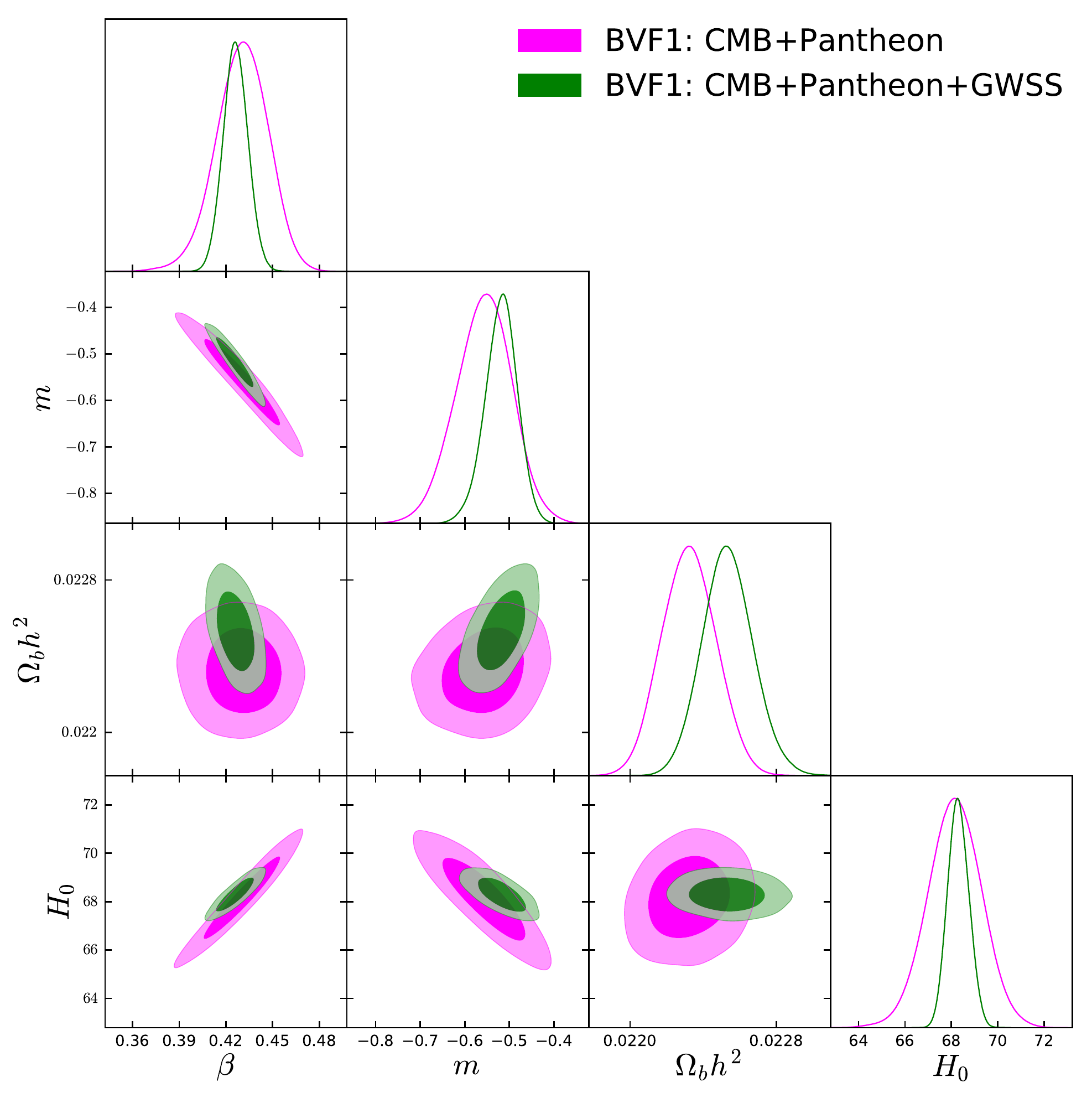}
\caption{For the BVF1 scenario we show the 1-dimensional posterior distribution of some model parameters and the 2-dimensional joint contours of the model parameters at 68\% and 95\% CL for CMB+Pantheon and CMB+Pantheon+GWSS.  }
\label{fig:bvf1-cmb+pan-vs-gw}
\end{figure*}
\begin{figure}
\centering
\includegraphics[width=0.45\textwidth]{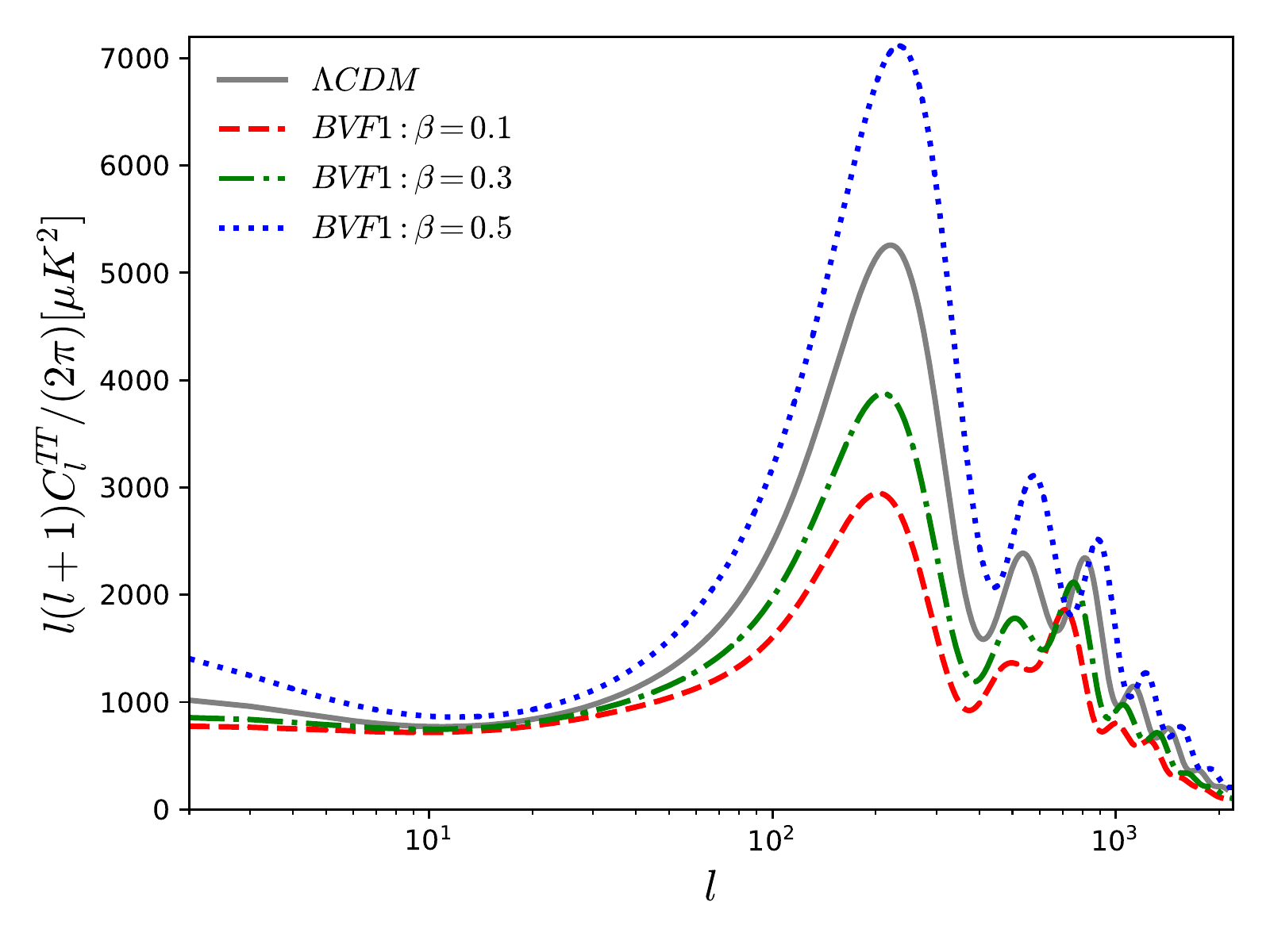}
\includegraphics[width=0.45\textwidth]{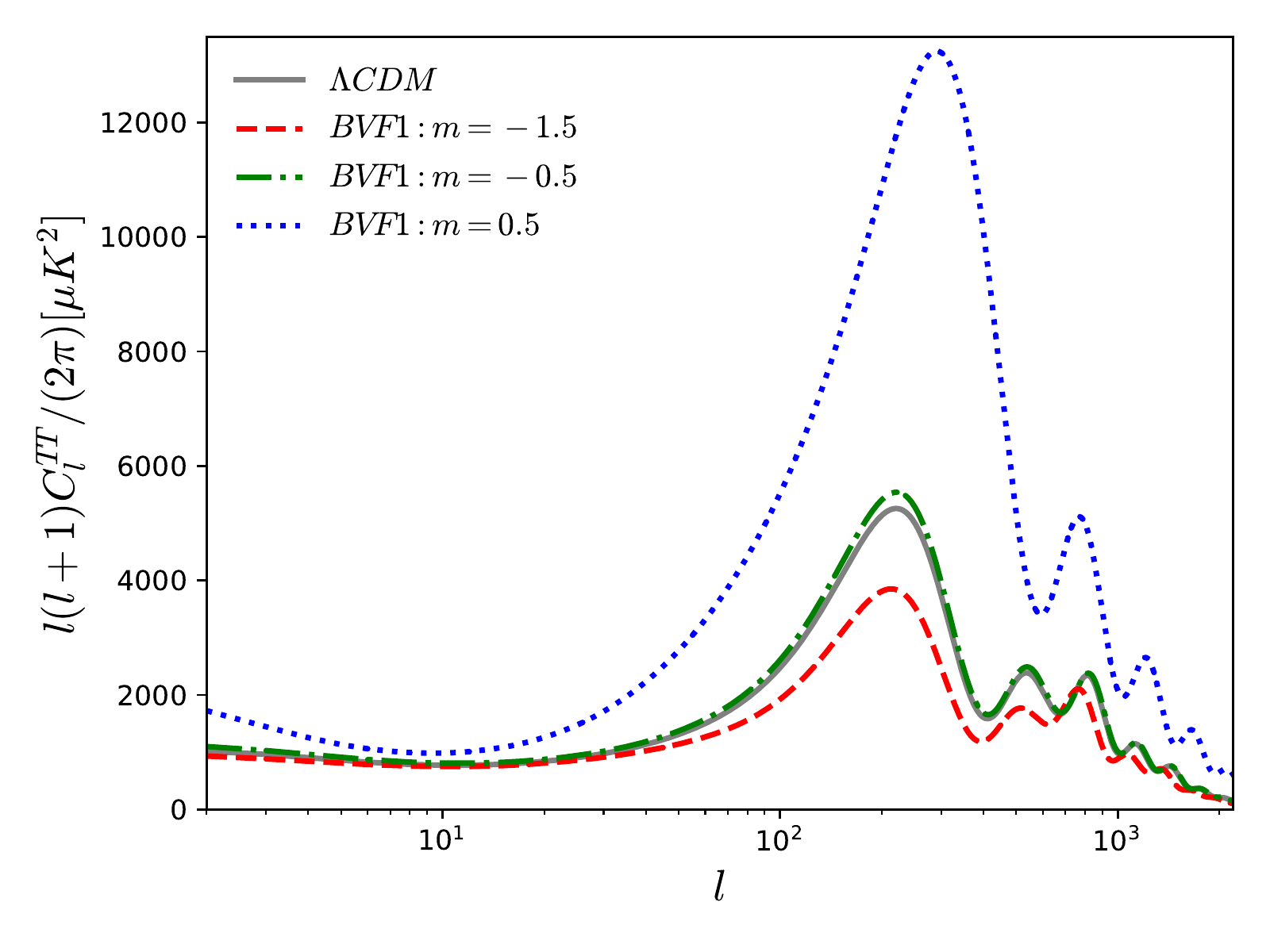}
\caption{ The CMB $C^{TT}_{l}$ power spectrum versus multipole moment $l$ using the best fits values obtained for the BVF1 model using the join data sets described, with three arbitrary $\beta$ and $m$ values.}
\label{fig-bvf1:cmb-TT}
\end{figure}
%%%%%%%%%%%%%%%%%%%%%%%%%%%%%%%%%%%%%%%%%%%%%%%%%%
%%%%%%%%%%%%%%%%%%%%%%%%%%%%%%%%%%%%%%%%%%%%%%%%%%%
\section{Observational constraints: Results and analysis}
\label{sec:method}

In this section we present the constraints on the bulk viscous scenarios considering CMB+Pantheon and CMB+Pantheon+GWSS. As we are interested to estimate the improvement of the cosmological parameters in presence of the GWSS measurements, and as the combined standard cosmological probes offer the most stringent constraints on the cosmological parameters, therefore, the inclusion of GWSS with the combined standard cosmological probes is reasonable.  
As mentioned earlier, the key common free parameters of BVF1 and BVF2 are $\beta$ and $m$, since $\beta \neq 0$ indicates the preference for a non-zero bulk viscosity and $m \neq 0$ indicates that the coefficient of the bulk viscosity is not constant in the redshift range considered. In the following subsections we discuss the constraints on these two scenarios in detail. 

%%%%%%%%%%%%%%%%%%%%%%%%%%%%%%%%%%%%%%%%%%%%%%%%%%%

\subsection{Constraints on the BVF1 scenario}
\label{sec-bvf1}

In Table~\ref{tab:results-BVF1} we have presented the constraints on the BVF1 scenario for CMB+Pantheon and CMB+Pantheon+GWSS. The latter dataset is aimed to understand the improvement expected from GWSS on the constraints from CMB+Pantheon. 
In Fig.~\ref{fig:bvf1-cmb+pan-vs-gw} we have compared these datasets graphically by showing the one dimensional marginalized distribution of some model parameters and the two dimensional joint contours. As discussed, this scenario has two main key parameters, namely, $\beta$, quantifying the existence of bulk viscosity in the cosmic sector, and $m$, which tells us whether the bulk viscosity will have a dynamical nature (corresponding to $m \neq 0$) or not.

Since for CMB+Pantheon, we find an evidence for a non-zero bulk viscosity in the cosmic sector at many standard deviations, i.e.~$\beta =  0.430_{-    0.016}^{+    0.017}$ at 68\% CL, this is further strengthen for CMB+Pantheon+GWSS, where $\beta = 0.4262_{-    0.0078}^{+    0.0079}$ at 68\% CL.\footnote{It is worthwhile to note here that the mean value of $\beta$ is not significantly changed when the GWSS data are included, because we built the simulated data using the best fit obtained 
from CMB+Pantheon. } One can clearly see that the inclusion of GWSS to CMB+Pantheon improves the error bars on $\beta$ by a factor of at least 2.  This reflects the constraining power of GWSS.  On the other hand, focusing on the parameter $m$, which quantifies the time evolution of the bulk viscosity, we see that $m$ remains non-zero at several standard deviations  for CMB+Pantheon, where the 68\% CL constraint on $m$ is $m = -0.557_{-  0.059}^{+    0.068}$, and becomes $m = -0.519_{-  0.035}^{+    0.038}$ for CMB+Pantheon+GWSS. 
From the constraints on $m$, one can clearly see that the uncertainty in $m$ is reduced by a factor of $\sim 1.7 - 1.8$ when the GWSS data are included with the combined dataset CMB+Pantheon. 
Concerning the Hubble constant, we find that $H_0$ assumes slightly higher values compared to the $\Lambda$CDM based Planck. Actually, we have $H_0 = 68.1_{-    1.1}^{+    1.2}$ at 68\% CL for CMB+Pantheon, while $H_0 = 68.30_{-    0.45}^{+0.46}$ at 68\% CL for CMB+Pantheon+GWSS, again improving the uncertainty in $H_0$ by a factor of $2.5$. This shows that the effects of GWSS are clearly visible through these parameters. In Fig.~\ref{fig:bvf1-cmb+pan-vs-gw}, one can compare the constraints on the model parameters obtained from CMB+Pantheon and CMB+Pantheon+GWSS.

Finally, through Fig.~\ref{fig-bvf1:cmb-TT} we examine how the model affects the CMB TT power spectrum for different values of the model parameters, $\beta$ and $m$  with respect to the standard $\Lambda$CDM scenario. In the upper panel of Fig.~\ref{fig-bvf1:cmb-TT} we depict the evolution in the CMB TT power spectrum for different values of $\beta$ while in the lower panel of Fig.~\ref{fig-bvf1:cmb-TT} we depict the evolution in the CMB TT power spectrum for different values of $m$. From both the graphs, we notice that as long as $\beta$ or $m$ increases, the model exhibits significant differences in the lower multipoles ($\ell \leq 10$). For $\ell \geq 10$, we observe that with the increasing values of $\beta$ and $m$,  the peaks of the CMB TT power spectrum increase significantly, particularly changing their mutual ratio. 

\begingroup                                           
%\squeezetable                                           
\begin{center}                                              
\begin{table*}                                           
\begin{tabular}{ccccccc}                                
\hline\hline                                               
Parameters & CMB+Pantheon & CMB+Pantheon+GWSS \\ \hline

$\Omega_b h^2$ & $    0.02241_{-    0.00016-    0.00032}^{+    0.00016+    0.00032}$  &   $    0.02238_{-    0.00016-    0.00031}^{+    0.00015+    0.00030}$ \\

$100\theta_{\rm MC}$  & $    1.02907_{-    0.00082-    0.00198}^{+    0.00111+    0.00180}$  & $    1.02921_{-    0.00040-    0.00079}^{+    0.00041+    0.00080}$ \\

$\tau$ & $    0.0516_{-    0.0072-    0.015}^{+    0.0074+    0.015}$  &  $    0.0521_{-    0.0078-    0.014}^{+    0.0071+    0.015}$ \\

$n_s$ &  $    0.9575_{-    0.0066-    0.012}^{+    0.0053+    0.012}$  &  $    0.9583_{-    0.0039-    0.0077}^{+    0.0038+    0.0075}$ \\

${\rm{ln}}(10^{10} A_s)$  & $    3.038_{-    0.017-    0.032}^{+    0.016+    0.032}$   &  $    3.040_{-    0.015-    0.031}^{+    0.015+    0.032}$ \\

$\beta$  & $    0.447_{-    0.022-    0.042}^{+    0.022+    0.042}$ & $    0.425_{-    0.016-    0.034}^{+    0.018+    0.032}$ \\

$m$  & $   -0.85_{-    0.19-    0.50}^{+    0.30+    0.46}$    & $   -0.683_{-    0.089-    0.19}^{+    0.099+    0.18}$ \\

$\gamma$  & $    0.9970_{-    0.0024-    0.0036}^{+    0.0015+    0.0042}$  & $    0.99757_{-    0.00058-    0.0011}^{+    0.00049+    0.0011}$ \\

$H_0$ & $   65.2_{-    2.6-    3.9}^{+    1.7+    4.4}$  &   $   64.91_{-    0.60-    1.2}^{+    0.59+    1.1}$ \\

\hline\hline                                         
\end{tabular}
\caption{We report the observational constraints on the BVF2 scenario at 68\% and 95\% CL for CMB+Pantheon and CMB+Pantheon+GWSS.}                           
\label{tab:results-BVF2}                                
\end{table*}                                        
\end{center}                                          
\endgroup   
\begin{figure*}
\centering
\includegraphics[width=0.7\textwidth]{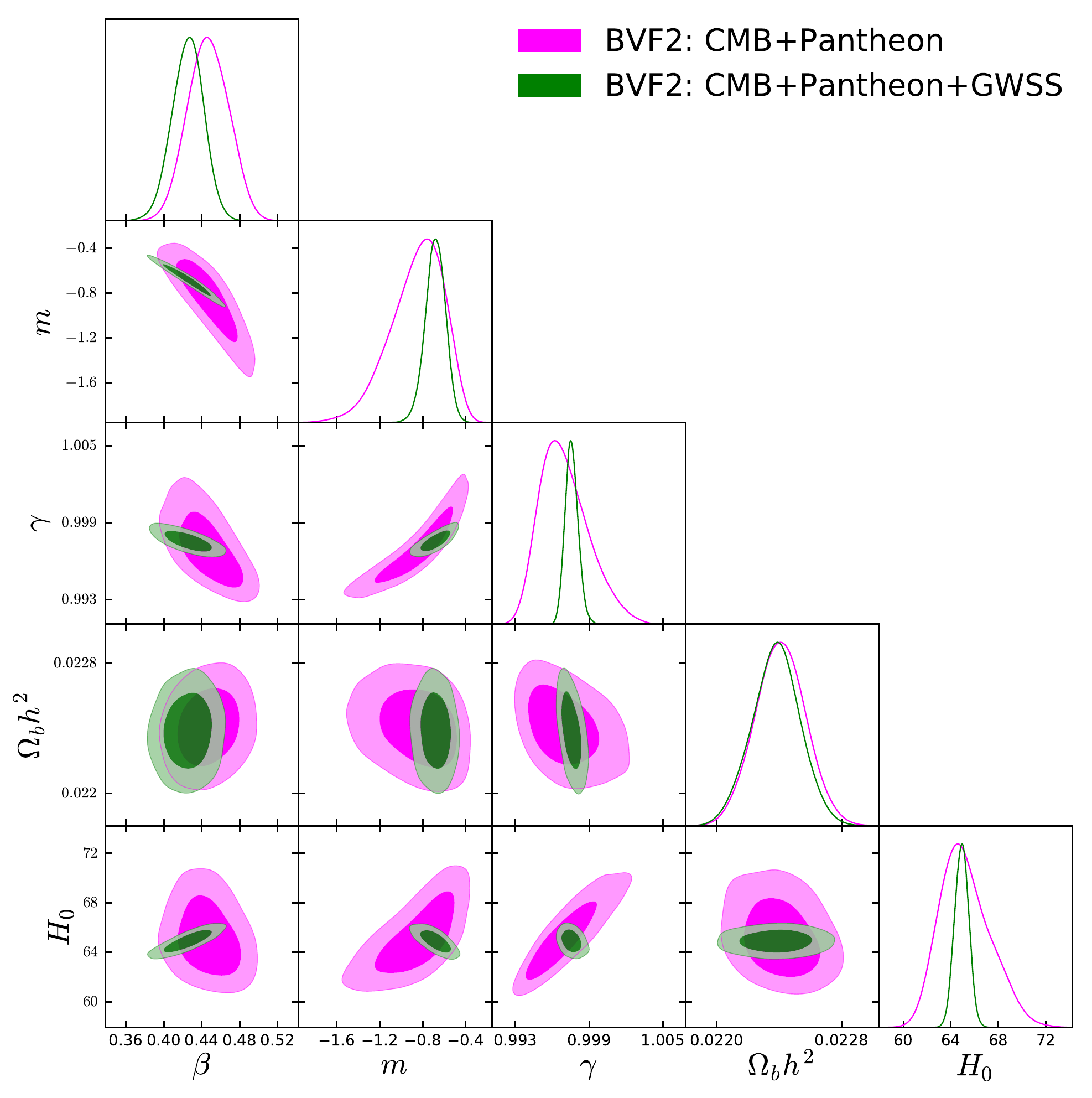}\caption{For the BVF2 scenario we show the 1-dimensional posterior distributions of some model parameters and the 2-dimensional joint contours of the model parameters at 68\% and 95\% CL for CMB+Pantheon and CMB+Pantheon+GWSS. }
\label{fig:bvf2-cmb+pan-vs-gw}
\end{figure*}
\begin{figure}
\centering
\includegraphics[width=0.45\textwidth]{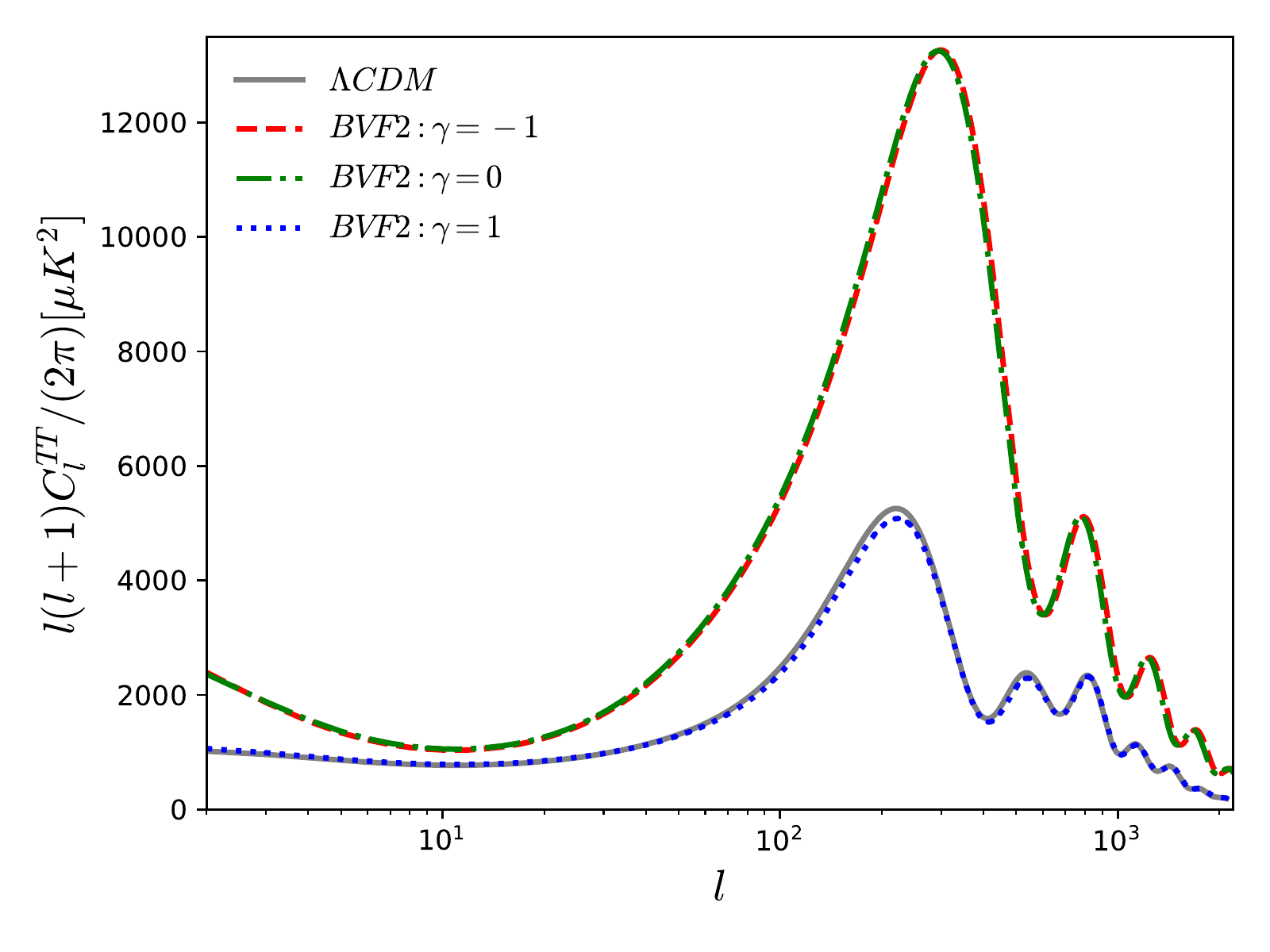}
\includegraphics[width=0.45\textwidth]{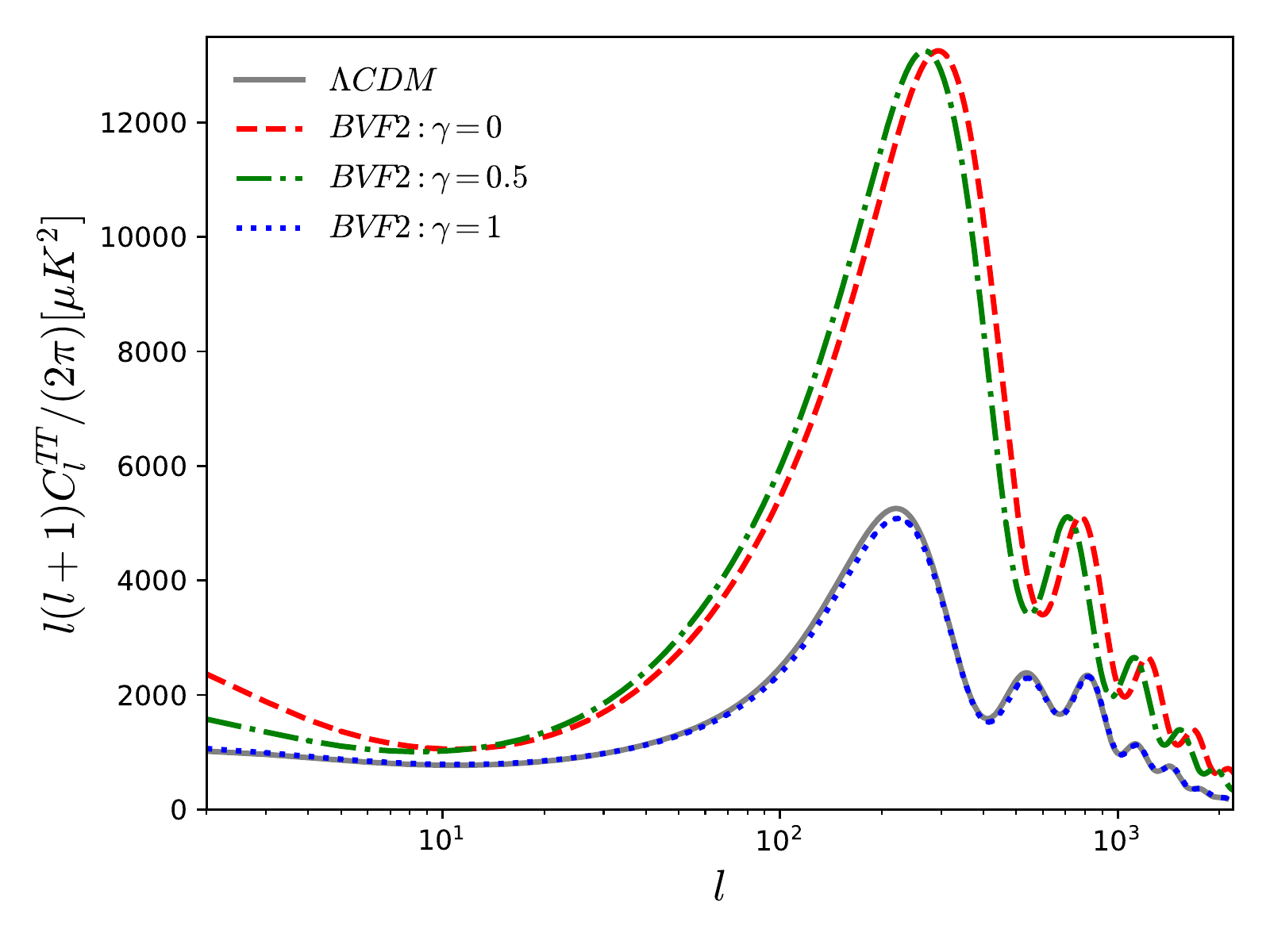}
    \caption{The CMB $C^{TT}_{l}$ power spectrum versus multipole moment $l$ using the best fits values obtained for each BVF2 models with three
    $\gamma$ values, respectively using the join data sets described.
    }
    \label{fig-bvf2:cmb-TT}
\end{figure}
%%%%%%%%%%%%%%%%%%%%%%%%%%%%%%%%%%%%%%%%%%%%%%%%%%%

\subsection{Constraints on the BVF2 scenario}
\label{sec-bvf2}

In Table~\ref{tab:results-BVF2} we present the constraints on the BVF2 scenario for both CMB+Pantheon and CMB+Pantheon+GWSS. And in Fig.~\ref{fig:bvf2-cmb+pan-vs-gw}, we compare the constraints from these datasets explicitly 
showing the one dimensional marginalized distribution of some model parameters and the two dimensional joint contours. As already discussed, this scenario has three main key parameters, namely, $\beta$, which quantifies the existence of bulk viscosity in the cosmic sector, $m$, which tells us whether the bulk viscosity enjoys a dynamical nature (corresponding to $m \neq 0$) or not, and finally, the parameter $\gamma$, which indicates the fluid which endows the bulk viscosity. We note that $\gamma  =1$ refers to the dust fluid endowing the bulk viscosity in which we are interested in, for which we recover the previous scenario BVF1.

For CMB+Pantheon, we find that $\beta \neq 0$ at several standard deviations yielding $\beta =  0.447\pm 0.022$ at 68\% CL which gives a clear indication of a non-zero bulk viscosity in the cosmic sector. When the GWSS are added to this combination, i.e. CMB+Pantheon+GWSS, the conclusion about $\beta$ does not change significantly ($\beta = 0.425_{- 0.016}^{+    0.018}$ at 68\% CL), indicating that for this scenario GWSS do not provide any additional constraining power on $\beta$. Looking at the dynamical nature of the bulk viscosity, we see that for CMB+Pantheon, $m$ remains nonzero at more than 2 standard deviations leading to $m = -0.85_{-    0.19}^{+    0.30}$ at 68\% CL. However, this evidence could be further strengthened by the inclusion of the GWSS data, that we forecast to be $  m = -0.683_{-0.089}^{+    0.099}$ at 68\% CL for CMB+Pantheon+GWSS, improving the error bars up to a factor of 3. Finally, focusing on the parameter $\gamma$ which directly connects with the nature of the cosmic fluid endowing the bulk viscosity, we can see that it is  consistent with $1$, which corresponds to a dust-like fluid, within 2 standard deviations for CMB+Pantheon ($\gamma = 0.9970^{+0.0015}_{-0.0024}$ at 68\% CL). Also for this parameter, the addition of the GWSS further improves the constraining power of a factor larger than $3$ to $4$, that we forecast to be $\gamma=0.99757^{+0.00049}_{-0.00058}$ at 68\% CL for CMB+Pantheon+GWSS. 
Therefore, with respect to the BVF1 case, where the inclusion of the forecasted GWSS was able to improve both $\beta$ and $m$, in this BVF2 scenario, the improvement of the constraining power is displayed only on $m$ and $\gamma$ but does not affect anymore $\beta$ significantly.

Furthermore, we find that for this scenario, the Hubble constant attains a very low value for CMB+Pantheon compared to Planck's estimation within the $\Lambda$CDM paradigm. 
We also note that $H_0$ is correlated to all three free parameters of this scenario, namely, $\beta$, $m$ and $\gamma$. With $\beta$ and $\gamma$, $H_0$ is positively correlated while with $m$, it has a strong  anti-correlation. For CMB+Pantheon, $H_0 = 65.2_{-    2.6}^{+    1.7}$ km/s/Mpc at 68\% CL and after the inclusion of GWSS  it becomes $H_0 = 64.91_{-0.60}^{+    0.59}$ km/s/Mpc at 68\% CL, reducing the uncertainty in $H_0$ by a factor of $\sim 4$.

Finally, in Fig.~\ref{fig-bvf2:cmb-TT}, we examine the CMB TT power spectrum for this bulk viscous scenario BVF2 considering  different values of the free parameter $\gamma$ with respect to the standard $\Lambda$CDM scenario.   As $\gamma$ lies in the region $[-3, 3]$ and the nature of the cosmic fluid characterized by its equation of state $p_u = (\gamma -1) \rho_u$ depends on the sign of $\gamma$, therefore, we have considered two separate plots, one where $\gamma $ is non-negative (i.e. $\gamma \geq  0$) and another plot where $\gamma$ allows both positive and negative values including $\gamma = 0$.  From both the panels of Fig.~\ref{fig-bvf2:cmb-TT}, we clearly see that any deviation from $\gamma = 1$ makes significant changes in the amplitude of the CMB TT power spectrum. 
In particular, we see that the peaks of the CMB TT spectrum significantly increases and shift towards higher multipoles for any value different from $\gamma =1$ at small scales, as well as the Integrated Sachs Wolfe (ISW) plateau at large scales. As $\gamma = 1$ indicates a cosmological constant-like fluid endowed with the bulk viscosity, therefore, for $\gamma = 1$, we replicate an equivalent behaviour of the $\Lambda$CDM scenario. 

%%%%%%%%%%%%%%%%%%%%%%%%%%%%%%%%%%%%%%%%%%%%%%%%%%%
\section{Conclusions}
\label{sec-conclu}

Although the $\Lambda$CDM cosmological model is extremely successful in describing a large span of astronomical observations, it cannot explain several theoretical and observational issues. This motivated the scientific community to construct a variety of cosmological proposals and testing them with the available astronomical data.  Among these cosmological models, in this article, we focus on the unified cosmological models allowing bulk viscosity in the background.  However, since these models do not recover the $\Lambda$CDM scenario as a special case, our only ability in distinguishing them, once the GWSS data will be available, will rely only on a Bayesian model comparison for a better fit of the cosmological observations, as done in Ref.~\cite{Yang:2019qza}.  
The unified cosmological scenarios endowed with bulk viscosity are appealing from two different perspectives: the first one is the concept of a unified picture of dark matter and dark energy, and the second is the inclusion of bulk viscosity into that unified picture. Effectively, the unified bulk viscous scenario is a generalized cosmic picture combining two distinct cosmological directions. According to a recent paper on the unified bulk viscous scenarios~\cite{Yang:2019qza}, current cosmological probes prefer a non-zero dynamical bulk viscosity in the dark sector at many standard deviations. So, in light of the current cosmological probes, unified bulk viscous cosmological scenarios are attractive.  In this line of thought,
what about the future of such unified bulk viscous scenarios? In this article we have focused on it with an answer.

Following Ref.~\cite{Yang:2019qza}, in this work we have explored these scenarios with the GWSS aiming to understand how the future distance measurements from GWSS may improve the constraints on such scenarios.  In order to proceed toward this confrontation, we have generated ${\cal O}(10^3)$ mock GWSS luminosity distance measurements matching the expected sensitivity of the Einstein Telescope and added these mock data to the current cosmological probes, namely CMB from Planck 2018 release\footnote{We mention that in the earlier work~\cite{Yang:2019qza}, CMB data from Planck 2015  were used to constrain the bulk viscous scenarios. } and SNIa Pantheon sample. We find that the inclusion of GWSS luminosity distance measurements together with the current cosmological probes makes the possible future evidence for new physics stronger, by reducing the uncertainty in the parameters in a significant way. This is a potential behaviour of the GWSS luminosity distance measurements since this makes the parameter much deterministic. Overall for both BVF1 and BVF2 scenarios, we find a very strong preference  of a non-zero time dependent bulk viscous coefficient (alternatively, the viscous nature of the unified dark fluid) at many standard deviations.

In conclusion, in the present paper we demonstrate that future GWSS distance measurements from the Einstein Telescope might be powerful to extract more information about the physics of the dark sector.  Therefore, based on the present results, we feel that it might just be a matter of time  before we convincingly detect the viscosity in the dark sector, if any.

%%%%%%%%%%%%%%%%%%%%%%%%%%%%%%%%%%%%%%%%%%%%%%%%%%%%%%%%%%%%%%%%%%%%%%%%%%%%%%%
\section{Acknowledgments}
The authors thank the referee for some important comments which helped us to improve the article considerably.   WY was supported by the National Natural Science Foundation of China under Grants No. 12175096 and No. 11705079, and Liaoning Revitalization Talents Program under Grant no. XLYC1907098.  SP acknowledges the financial support from  the Department of Science and Technology (DST), Govt. of India, under the Scheme
``Fund for Improvement of S\&T Infrastructure (FIST)'' [File No. SR/FST/MS-I/2019/41].
EDV is supported by a Royal Society Dorothy Hodgkin Research Fellowship.  
CE-R is supported by the Royal Astronomical Society as FRAS 10147 and by PAPIIT UNAM Project TA100122.  This article/publication is based upon work from COST Action CA21136 Addressing observational tensions in cosmology with systematics and fundamental physics (CosmoVerse) supported by COST (European Cooperation in Science and Technology).
AP was supported in part by the National Research Foundation of South Africa (Grant Number 131604). Also AP thanks the support of Vicerrector\'ia de Investigaci\'on y Desarrollo Tecnol\'ogico (Vridt) at Universidad Cat\'olica del Norte through N\'ucleo de Investigaci\'on Geometr\'ia Diferencial y Aplicaciones, Resoluci\'on Vridt  No - 096/2022.

%%%%%%%%%%%%%%%%%%%%%%%%%%%%%%%%%%%%%%%%%%%%%%%%%%%%%%%%%%%%%%%%%%%%%%%%%%%%%%
%\section*{Data Availability}
%In this work we have used  CMB power spectra from Planck 2018, Pantheon catalogue of Supernovae Type Ia and the mock GWSS data from the Einstein Telescope.   The first two datasets are publicly available while the last dataset is not available publicly.  The mock GWSS dataset might be shared upon reasonable request. 
%%%%%%%%%%%%%%%%%%%%%%%%%%%%%%%%%%%%%%%%%%%%%%%%%%%

\bibliography{biblio}

\end{document}